\documentclass[12pt,preprint]{aastex}
\begin{document}
\title{The Correlation Dimension of Young Stars in Dwarf Galaxies}
\author{Mary Crone Odekon}
\affil{Skidmore College, Saratoga Springs, NY 12866, USA}
\email{mcrone@skidmore.edu}

\begin{abstract}
We present the correlation dimension of resolved young stars in four actively star-forming dwarf galaxies that are sufficiently resolved and transparent to be modeled as projections of three-dimensional point distributions.  We use data in the Hubble Space Telescope archive; photometry for one of them, UGCA~292, is presented here for the first time.  We find that there are statistically distinguishable differences in the nature of stellar clustering among the sample galaxies.  The young stars of VII Zw 403, the brightest galaxy in the sample, have the highest value for the correlation dimension and also the most dramatic decrease with logarithmic scale, falling from $1.68\pm0.14$ to $0.10\pm0.05$ over less than a factor of ten in $r$.  This decrease is consistent with the edge effect produced by a projected Poisson distribution within a 2:2:1 ellipsoid.  The young stars in UGC~4483, the faintest galaxy in the sample, exhibit very different behavior, with a constant value of about 0.5 over this same range in $r$, extending nearly to the edge of the distribution.  This behavior may indicate either a scale-free distribution with an unusually low correlation dimension, or a two-component (not scale-free) combination of cluster and field stars.  

\end{abstract}


\section{Introduction}

Two striking and controversial features in the distribution of gas and young stars in galaxies are their scale-free structure and their universality across different environments.  Approximate forms of self-similarity are manifest through a variety of analysis techniques, including intensity power spectra, perimeter-area scaling, velocity-size relations, and autocorrelation functions, on scales ranging from several AU all the way up to entire galaxies. (See Elmegreen \& Scalo 2004 for a review).   One explanation is that scale-free structure is linked to the turbulent transfer of energy among scales, but the importance of physical effects like gas compressibility, magnetohydrodynamics, gravitational fragmentation, and energy output from massive stars is not clear.  The identification of special scales where power laws change, and the identification of differences between the structure of gas and that of stars, may signal the regimes dominated by different physical processes (e.g. Larson 1995). 

Several authors specifically emphasize the universality of scaling laws across different environments.  Heyer \& Brunt (2004), for example, find similar stucture functions in 27 molecular clouds, and suggest that star formation is therefore dominated by a common mechanism like converging turbulent flows.  Willett, Elmegreen, \& Hunter (2005) emphasize the similarity in the power spectra of light in star forming galaxies over a large range of sizes.  Another recent result supporting universality in stellar clustering is the similarity in clustering properties of OB stars in the Small Magellanic Cloud in clusters and the field (Oey et al. 2004).  

Previous studies of the projected distribution of starlight on galaxy-wide scales include Elmegreen \& Elmegreen (2001), who conclude that the starlight in spiral galaxies is consistent with a fractal dimension of 1.35 (a typical value for gas clouds), and Parodi \& Bingelli (2003) who find that star-forming complexes in dwarf irregular galaxies have cluster dimensions between 1.3 and 2.  However, several difficulties complicate the interpretation and comparison of these studies.  One is blending:  large-scale clumps tend to hide the smaller-scale structure within them ---  indeed, the more highly clustered the distribution, the more likely that small-scale structure will be hidden within large-scale structure.  A related, but distinct, problem is varying opacity.  A region need not be visibly crowded to obscure starlight; the culprit may be dark dust.  Another difficulty for observations based on optical light from stars is that clustering properties depend on the relative contributions from stars of different ages, older stars having had more time to relax into a smooth distribution. 

In this paper, we minimize these problems by selecting galaxies that are sufficiently resolved and instrinsically transparent to be modeled as projected three-dimensional distributions.  In other words, blending and extinction are minimal for the stars and physical scales under consideration.  In addition, we use resolved stars with ages we can estimate from their positions on a color-magnitude diagram.  These constraints greatly simplify the quantification of uncertainties and comparison with models. 

\section{Data and Photometry} 

Our choice of sample galaxies is driven by the goal of modeling the distribution of young stars as the projection of a three-dimensional distribution of points.  We use very faint, low-metallicity dwarfs, for which extinction by dust is low and the line-of-sight depth is small (Table 1).  The absolute magnitudes $M_B$ for our sample galaxies range from $-11.43$ to $-14.30$.  For comparison, the 114 galaxies in the Palomar/Las Campanas Imaging Atlas of Blue Compact Dwarf Galaxies have a considerably brighter typical magnitude of $M_B = -16.1$ mag (Gil de Paz, Madore, \& Pevunova 2003).   Additional advantages of these galaxies are their small sizes on the sky and relatively high galactic latitudes, resulting in little contamination by galactic extinction or foreground stars.   

The four galaxies in our sample have deep Wide Field Planetary Camera 2 (WFPC2) images in the Hubble Space Telescope Archive.  Single-star photometry for three of them has been published previously, while the photometry for UGCA~292 is published here for the first time.  For this study, we performed the photometry for all four galaxies beginning with the original images, in order to assess completeness and blending.  

We examined archival data for several other dwarf galaxies and found them to be insufficient for our purposes because of shorter exposure times (snapshot survey galaxies), too much crowding and extinction (brighter galaxies like NGC~1549 and more distant galaxies like I~Zw~18), or limited coverage of the star forming region (local group galaxies).  Some of these galaxies are prime targets for future deep high-resolution images, and, indeed, observations of some of them are currently underway with the Advanced Camera for Surveys. 

Upon inspection, the images of the sample galaxies appear nearly transparent in the sense that distant background galaxies show through (Fig.~1;  see also Fig.~2 in Izotov \& Thuan 2002 and, for an especially obvious example, Fig.~1 in Crone et al. 2002).  Further evidence for low extinction comes from the color-color diagrams for  VII~Zw~403 and UGCA~290, which show no evidence for internal reddening (Crone et al. 2002).  There are a few small clusters, some with intense nebular emission, that are not transparent.  For example, Izotov \& Thuan find a patchy structure in the $V-I$ subtracted image of the emission nebula surrounding the dense cluster in UGC~4483, that they interpret as differential extinction.  As discussed below, these small-scale instances of crowding and extinction are quantifiable, and do not preclude calculating the correlation dimension on sufficiently large scales.   

Data reduction is essentially the same for each of the four galaxies. Here, we describe the specific process for UGCA~292;  details for the other galaxies, including errors and completeness, can be found in the references in Table 1.  The observations for UGCA~292 use nine dither positions.  We combined exposures at the same pointings using the CRREJ task within the STSDAS software package, and combined the different pointings using DRIZZLE onto a $1600\times1600$ grid.  After masking out obvious background galaxies and foreground stars, we conducted single star photometry with DAOPHOT using zero points from the 1997 May SYNPHOT tables.  To assess the completeness and accuracy of the photometry, we used ADDSTAR to superpose artificial stars of known magnitude on the inner 60 arcsecond by 40 arcsecond region of the galaxy (the rectangular region in Fig. 1).  To avoid artifical overcrowding, we repeatedly performed the photometry with only ten additional stars at a time, for a total of 9000 stars over the magnitude range $21-30$ in each filter.   The percentage of artificial stars recovered, as well as the difference between the input magnitude and the recovered magnitude, are in Fig. 2.  

We corrected for the small foreground extinction according to the values in Schlegel, Finkbeiner, \& Davis (1998), and transformed the F555W and F814W magnitudes into V and I following Holtzman et al. (1995).   The resulting CMD (Fig. 3) shows features characteristic of star forming dwarf galaxies:  a plume of very blue ($V-I<0$) stars that generally includes blue supergiants, very blue main sequence stars, and blue Helium-burning stars; and a parallel plume of slightly red ($V-I\sim1$) stars that generally includes red supergiants, giants, and asymptotic giant branch stars.

We estimate the distance to UGCA~292 using the tip of the red giant branch (TRGB), which occurs consistently at $M_I\sim-4.0$ for low metallicity stars (Lee, Freedman, \& Madore 1993).  The horizontal line in Fig.~3 indicates the most likely magnitude of the TRGB, based on the rise in the luminosity function of red ($V-I>0.5$) stars at $I\sim 24.5\pm0.2$;  this edge in the luminosity function appears as a positive signal from a Sobel edge detector, shown as the dotted histogram in Fig.~3.  Note the smaller rises at $I\sim 24.2$ and $I\sim 25.0$.  Dwarf galaxies often exhibit a rise in the red luminosity function caused by asymptotic giant branch (AGB) stars that are slightly brighter than the TRGB, but is often clear from the CMD morphology that the larger peak is indeed the TRGB (see Fig. 4 and Schulte-Ladbeck, Crone, \& Hopp 1998 for the case of VII~Zw~403, which shows a red spray of AGB stars above the TRGB).  In the case of UGCA~292, we take the largest peak to be the TRGB, with a statistical uncertainty of 0.2 mag, keeping in mind the possibility that the peak at 24.2 might be the actual TRGB.  It is even possible that the rise at $I\sim 25.0$ is TRGB, but this would imply a strangely prominent AGB population;  population synthesis modeling with very low metallicity evolutionary tracks might provide a solid test for this possibility. 

Photometric errors for the TRGB stars in UGCA~292 are small (nearly all less than 0.1 mag; see Fig. 2) and overall completeness is nearly 100\%.  This region of the CMD is also fairly well sampled, with over 300 red stars in the one-magnitude interval below the TRGB.   There is, however, an uncertainty caused by the extremely low metallicity of this galaxy.  As expected from its low nebular abundance (12+log(O/H)=7.32, van Zee \& Haynes 2006), the red giants are abnormally blue.  Following Salaris \& Cassisi (1997), we can estimate the metallicity of red giant stars from the $(V-I)_o$ color of the RGB half a magnitude fainter than the TRGB.  For UGCA~292, this color is in the range $1.0-1.1$ mag, corresponding to a metallicity just outside the range of their TRGB calibration ($1.2 < V-I < 2.0$ mag, corresponding to the metallicity range $-2.23 < [M/H] < -0.57$).  We will still use the usual value for the TRGB magnitude, $M_I = -4.0 \pm 0.1$, because it holds over this entire range of metallicity.  

Comparing our tip determination to the absolute TRGB magnitude, we find a distance modulus of $28.5 \pm 0.24$ mag, corresponding to a distance of $5.0 \pm 0.4$ Mpc.  If the rise at 24.2 is actually the TRGB, then UGCA~292 is a bit closer, at 4.0~Mpc.   Either way, our distance estimate is larger than the 3.1~Mpc obtained using Virgocentric infall models (van Zee \& Haynes 2006). 

In general, TRGB distance determinations for the other galaxies are more straightforward than that for UGCA~292, thanks to more obvious tip locations on the CMD and less extreme metallicity. For details, see the references in Table 1.  We find distance moduli of $29.10 \pm 0.12$ for UGCA~290, $28.23 \pm 0.10$ for VII~Zw~403, and $27.7 \pm 0.15$ for UGC~4483.  Our distance determinations for UGCA~290 and VII~Zw~403 are identical to those in the references in Table 1.   Our distance determination for UGC~4483, $3.5 \pm 0.2$ Mpc, is consistent with Izotov \& Thuan (2002), who find $3.4 \pm 0.2$ Mpc from the same data set.  Fig. 4 shows the V, I color-magnitude diagrams (CMDs) in terms of absolute magnitude for all four galaxies.  

The selection criteria for young stars balance several considerations.  There should be enough stars to produce statistically useful results.  Photometry for the selected stars should be complete or nearly complete, requiring sufficiently bright stars.  Finally, the stars should be sufficiently young that clusters on the scales under consideration have not dissipated, and we must be able to determine their ages from positions on the CMD.  

For age determination, we use the results of Crone et al. (2002;  see in particular their Fig. 5), who  modeled the star formation histories of UGCA~290 and VII~Zw~403 with the Bologna synthetic code (Greggio et al. 1998).  Note that synthetic modeling improves upon simple comparison with isochrones because it quantitatively estimates the degree to which each area of the CMD includes populations with a mix of different ages.  They found that two sections of the CMD are limited to stars approximately in the range 0 to 20 Myr, with a few as old as 30 Myr:  the brightest main sequence stars ($V-I<-0.2$, and $M_I<-4$) and the brightest supergiants ($M_I<-7.5$).  For our sample galaxies, the number of stars in this population ranges from 32 for UGC~4483 to 125 for UGCA~290.  We also consider a larger population including supergiants down to $M_I = -6$, corresponding to stellar ages up to about 100 Myr.  For our sample galaxies, the number of stars in this population ranges from 47 to 164.  Finally, for comparison, we also calculate the correlation dimension for all the resolved stars in each galaxy, including red giant stars billions of years old.     

Dissipation timescales for clusters on scales of tens of parsecs are likely to be hundreds of millions of years, so we do not expect the youngest stars  to be significantly affected by dynamical relaxation on these scales.  The crossing time $\tau_c$ for a 10 pc cluster with velocity dispersion 1 km s$^{-1}$ (typical for nearby clusters) is 10 Myr.  According to the analytical approximation of Binney \& Tremaine (1987), the dynamical relaxation time is $\tau_{rel} \sim (0.1N/\rm{ln}N)\tau_c$ , so a cluster with only $10^3$ stars has $\tau_{rel}$ as large as $\sim 300$~Myr.  For the scales considered in this paper, both $N$ and cluster sizes are higher, producing still longer relaxation timescales.  Velocity dispersions on these scales are very unlikely to be high enough to counter these effects: galaxy-wide gas velocity dispersions are only about 10 km s$^{-1}$ in galaxies this small (e.g. van Zee, Salzer, \& Skillman 2001).  It is also possible to estimate relaxation timescales through N-body simulations.  Goodwin \& Whitworth (2004) find that the structure in initially fractal star clusters is erased in one to several crossing times, depending on the initial velocity dispersion.  In addition to these theoretical considerations, it should also be noted that many nearby galactic clusters with ages of hundreds of millions of years and sizes on the order of 10 pc are known to exist as discernable entities (e.g. Chen, Chen, \& Shu 2004).  

Fig. 5 shows the spatial distribution of the stars in the second age group ($0-100$~Myr) for each galaxy, compared with R-band and B-band images from the online database of the Palomar/Las Campanas Imaging Atlas of Blue Compact Dwarf Galaxies.  (For UGCA~292, which is not part of the Las Campanas Atlas, images are from the lower-resolution POSS-II survey.)  The R-band isophotes approximate the extent of the older, dynamically relaxed stellar population, while bright clumps in the B band represent the extent of star-forming complexes.  Comparison of these two populations shows that the WFPC2 field of view succeeds in covering most or all of the star-forming region in each galaxy.  Note that, unlike the elliptical distribution of older stars, there is not generally a concentrated center in the distribution of young stars. 

These images also illustrate the range of galaxy morphologies in our sample.  UGCA~290, with only a small elliptical distribution of older stars, is intermediate between a blue compact dwarf (BCD) and a dwarf irregular.  VII~Zw~403, meanwhile, is often cited as a classic type nE BCD, exhibiting central star formation within a large background of older stars (Isotov, Thuan, \& Lipovetsky 1997).  An interesting connection between these two galaxies is that despite their difference in morphology, the CMDs for their young stars (stars with $M_I < -4, M_V < -3.7$) are statistically indistinguishable (Crone et al. 2002).  UGCA~292 is a dwarf irregular with the intriguing properties of especially low metallicity and especially high neutral hydrogen gas content $M_H/L_B = 6.9$ (van Zee 2000), suggesting that it is in a very early stage of development.  Finally, UGC~4483 is a cometary-shaped galaxy sometimes cited as a BCD.  However, based on its low peak surface brightness, Gil de Paz, Madore, and Pevunova (2003) conclude that it ``should certainly not" be classified as a BCD but as a dwarf irregular. Many of the young stars in UGC~4483 are limited to a compact star-forming complex near its northern edge.  

As illustrated in Fig. 2 and the references in Table 1, artificial star tests that scatter stars at random over the face of each galaxy indicate nearly complete photometry for the young stellar populations we consider.  However, if these stars are indeed clustered, stars may still be lost at a significant rate in a few small, crowded regions.  In order to assess this issue, we determined whether unrecovered artificial stars are limited to small, obviously crowded regions, and if so, what the sizes of those regions are.  We emphasize that this method relies on using stars for which photometry across most of the galaxy is nearly complete, and for which the few missing stars can be clearly identified with obviously crowded clumps.  Otherwise, it is not possible to know the actual underlying distribution except through model-dependent consistency tests --- the usual blending problem encountered in lower-resolution images.  

For each galaxy, we examined the locations of lost artificial stars for three populations:  the main sequence at $-4.75 < M_I < -3.25$, supergiants at $-6.25 < M_I < -5.75$, and bright supergiants at $-8.25 < M_I < -7.75$.  In other words, we considered the faint end of each region on the CMD that we use to select young stars.  For all the galaxies, the very small percentage (0-1\%) of lost bright supergiants are within two pixels (2 -- 3 pc) of the center of very bright neighboring stars.  The results are the same for the fainter supergiants, except in the case of VII~Zw~403, where the completeness rate is down to 75\% within a bright knot 20 pc across.  For the main sequence stars, as well, the only stars lost are within obvious dense clumps, but these clumps are larger:  40 pc (UGCA~290), 50 pc (VII~Zw~403), 23 pc (UGC~4483), and 36 pc (UGCA~292).  In each of these cases, completeness for artificial stars randomly scattered \textit{within the clumps} is $50\%-95\%$. Because the effective resolution is worse for the main sequence stars, we consider the supergiants separately, as well as augmented by the young main sequence stars, in the analysis that follows. 

\section{The Correlation Dimension: Definitions}  

A self-similar distribution of points can be partially characterized by its correlation dimension $d_c$: 
\[ N(r)  \propto r^{d_c} \]
where the correlation integral $N(r)$ is the average number of particles in a region of radius $r$.  For a random Poisson distribution in three-dimensional space, for example, $N$ simply scales as the volume, yielding a correlation dimension $d_c = 3$;  the Poisson distributon fully samples the three-dimensional space.  Various physical processes correspond to other predictions for $d_c$.  The isothermal surfaces for simple Gauss Kolmogorov incompressible turbulence follow a fractal pattern with a dimension of 2.67 (Mandelbrot 1983), while other scaling laws obtain for turbulence models that include other, perhaps more realistic, physical processes (Elmegreen \& Scalo 2004).  

Projection onto the two-dimensional plane of the sky reduces the expected value of $d_c$.  Analytically, the projection of a mathematical fractal with dimension $d$ into two dimensions follows the rule $d_{proj} = d$ for $d < 2$, while $d_{proj} = 2$ for $d \geq 2$ (Falconer 1990).   S\'anchez, Alfaro, \& P\'erez (2005) tested this behavior for projected fractal clouds sampled by a finite number of points, and found that the correlation dimension roughly followed this behavior for $d$ less than about 1.5 and greater than about 1.9.  At intermediate values, $d_{proj}$ was less than the analytic description --- for example, $d=2$ produced $d_{proj} = 1.7$. Note that these results apply specifically to projections rather than slices;  a two-dimensional slice, on the other hand, yields a dimension less by 1.

Regardless of whether the distribution of stars is actually fractal, the correlation dimension as a function of scale $d_c(r)$, like any correlation function, provides a simple way to characterize a distribution that is hierarchically clustered.  The correlation dimension, in particular, is used extensively to characterize time series data (Grassberger, Schreiber, \& Schaffrath 1991).  Its use for spatial distributions in astronomy includes application to young stars in Taurus (about 1.4, Larson 1995), water masers in galactic star-forming regions (between about 0.2 and 1.0, Strelnitski et al. 2002) and star forming complexes in dwarf galaxies (between about 1.2 and 1.8, Parodi \& Binggeli 2004.) 

We calculated $N(r)$ by counting the average number of stars within a distance $r$ of each existing star.  This is equivalent to the Grassberger \& Procaccia (1983) pair-counting method, multiplied by the total number of particles.  We then found $d_c$ by fitting a power-law function to $N(r)$ within logarithmic bins in $r$.  

Fig. 6 illustrates the ability of this technique to recover the dimension of a simple finite fractal randomly sampled by points.  We use the Sierpinski triangle, a well-known fractal in the shape of a triangle composed of three smaller triangles half as large as the first, each of which is composed of three trianges half as large, and so on.  This construction produces the fractal dimension log3/log2 $\sim 1.585$.  As shown on the left panels in Fig. 6, our ability to recover the fractal dimension is limited by both the finite length of the fractal and the limited number of points.  On large scales, edge effects become important, flattening $d_c$ to zero when $r$ becomes large enough to include all the particles.  Regardless of the number of particles, the finite length of the fractal limits the scales over which $d=1.585$ is recovered to those less than about a quarter of the total length.   On small scales, meanwhile, the ability to recover $d$ is limited by the number of particles available to sample it.  For this particular fractal, it requires at least 50 particles to find 1.585 over an order of magnitude in $r$, and 500 particles to find 1.585 over two orders of magnitude.  

Limited spatial resolution is an additional concern.  The panels on the right of Fig. 6 illustrate this problem, again using a Sierpinski triangle of length 1000, but this time with sharp resolution cutoffs at 1, 5, 10, and 20.  Note that the steepening of $N(r)$ on small scales extends to scales considerably larger than the cutoff, about five times $r_c$.  These sharp cutoffs more closely  model the formal resolution scale of the images (e.g. the full width at half maximum of the point spread function), than the dense clumps, which are complete to better than 50\%, but are a warning that resolution effects may creep in to length scales as large as five times the size of dense clumps.

The models in Fig. 6 do not address the combination of edge and projection effects that apply to a finite distribution in three dimensions.  In particular, the front and back edge of a three-dimensional distribution ``appear" throughout an entire two-dimensional projection.  A simple model that includes these effects is a random Poisson distribution bounded by a three-dimensional ellipsoid. We consider four specific models within this class:  an edge-on oblate ellipsoid with axial ratio 1:2:2, a side-view prolate ellipsoid with axial ratio 1:1:2, a sphere, and a face-on disk (equivalent to an end-on cylinder).  As expected, the pure Poisson dimension of two is obtained for small scales, but at larger scales the dimension gradually decreases to zero (Fig. 7).  The edge effect is smallest for the two-dimensional disk, where the Poisson dimension $d_c = 2.0$ is recovered for scales up to about half of the semimajor axis.  The problem is greatest for the ellipsoids, where projection carries the edge effect to scales as small as $r=250$, only one fourth of their semimajor axis.   

A further refinement of these projected Poisson models includes a decreasing volume number density.  Fig. 8 shows results when the density decreases exponentially with ellipsoidal isodensity contours, such that the density falls by a factor of $e^{10}$ when the distance from the center along the major axis is 1000.  The behavior of the cluster dimension is very similar to the constant density models over the scales where $d_c$ falls from 2.0 to about 0.5;  for example, in both cases the oblate ellipsoids fall from 1.8 to 0.5 over a factor of five (a difference of 0.7 in log$r$).  On larger scales, $d_c$ decreases more gradually for the exponential models, reflecting the smearing out of the edge.  

These models span a wide range of plausible shapes and orientations.  For example, an end-on prolate ellipsoid has projection effects between those of a sphere and an end-on cylinder.  Similarly, the behavior of a partially inclined, flat oblate ellipsoid is between that of the disk and the edge-on 2:2:1 ellipsoid.  Distributions that are more extreme, such as a very flat, edge-on ellipsoid are not consistent with the observed projected distribution of stars (Fig. 5).  

\section{The Correlation Dimension of the Sample Galaxies}  

Qualitatively, the correlation integrals for the actual stellar populations show the same behavior as the models, with approximate power laws that flatten on larger scales and steepen on smaller scales (Fig. 9).  It is also apparent that the young population of UGC~4483 exhibits a considerably shallower slope, and that there are slight differences between the slopes for different ages and between young main sequence and young supergiants.   But it is not clear from Fig. 9 alone whether these differences are a statistical effect.  In particular, the young population in UGC~4483 is fairly small, at just 32 stars.  (Note that the total number of stars in each population is indicated by the maximumum value of $N(r)$, and the spatial extent of the population by the value of $r$ where the slope flattens to zero.)  In order to make quantitative comparisons, we fit $d_c$ within logarithmic bins in scale, and estimate uncertainties via bootstrap resampling.  More specificially, we create a set of fifty random realizations of each galaxy by repeatedly selecting from the original distribution of stars, fit $d_c$ in logarithmic bins of $r$, and use the standard deviation in these values as our measure of uncertainty.  

Fig. 10 illustrates the usefulness of inspecting the full behavior of $d_c(r)$ rather than a single value.  The correlation dimension for the young stars in VII~Zw~403, in particular, is not resolved over a wide enough range of scales to reveal a constant power-law, dropping from $1.7 \pm 0.1$  to nearly zero over less than an order of magnitude in scale.  The calculation of a single number for $d_c$ over these scales would be highly dependent on the range of scales chosen.  For other populations, the correlation dimension changes more slowly.  The youngest population in UGCA~290, for example, shifts from about 1.6 to 1.0 over about an order of magnitude in scale.  Even more extreme, the young stars of UGC~4483 show a consistently low value for $d_c(r)$, at about 0.5.  This galaxy has a dense young cluster on one side; perhaps this behavior reflects a non-scale free combination of two distinct components:  a loose distribution of field stars and the compact cluster.  

As illustrated above, a sharp resolution cutoff steepens $d_c(r)$ on scales up to five times larger than the cutoff scale.  For the supergiant populations, this does not pose a problem over the range in Fig. 9;  their photometry is complete down to just a few parsecs.   The photometry for the bright main sequence stars, on the other hand, is only 50-95\% complete within a few specific clumps a few tens of parsecs across.  A cautious estimate of largest scales that might be affected is five times the clump size:  $\rm{log}r=2.3$ for UGCA~290, 2.4 for VII~Zw~403, 2.25 for UGCA~292, and 2.1 for UGC~4483.  Because the clump size is certainly not a sharp cutoff, this effect should be small for the scales in Fig. 10.  The implication for interpreting the results in Fig. 10 is that the first two or three data points in bold for the $0-30$ Myr and $0-100$ Myr populations may be slightly too high.  In fact, the supergiant and main sequence populations give statistically equivalent results.  

For comparison, we also calculate $d_c(r)$ for the entire resolved stellar population in each galaxy, a population that is both incomplete and heterogeneous in age.   As expected for older stars with time to dynamically relax, $d_c(r)$ is very close to 2.0 on small scales and decreases in a manner similar to the exponential ellipsoid. 

\section{Summary and Discussion} 

Using high-resolution data for four very small star-forming galaxies, we are able to calculate the correlation dimension as a function of scale with quantified uncertainties.   We find that there are statistically distinguishable differences in the nature of stellar clustering among the sample galaxies.  The young stars of VII Zw 403, the brightest galaxy in the sample, show the highest dimension and also the most dramatic decrease with logarithmic scale, falling from $1.68\pm0.14$ to $0.10\pm0.05$ over less than a factor of ten in $r$.  This decrease is consistent with the edge effect produced by a random distribution of points bounded by a 2:2:1 ellipsoid.  The young stars in UGC~4483, the faintest galaxy in the sample, exhibit very different behavior, with a constant value of about $0.5$ over this same range in $r$, extending nearly to the edge of the distribution.  This low, flat $d_c(r)$ is not consistent with a random distribution bounded by any ellipsoidal shape that fits the stellar distribution on the sky. 

The errors from bootstrap resampling are small enough to support the significance of a small dimension for UGC~4483.  Further support that it is not a simple statistical effect comes from Fig.~6:  while small populations produce larger uncertainties, there is no trend toward smaller dimension.  As an additional test, we randomly subsampled the stars in UGC~290 to match the total number of stars in UGC~4483.  The results were consistent with those for the full UGC~290 population, and not with UGC~4483.   

While our sample is small, it is worth noting that the brightest galaxy produces the highest dimension and the faintest galaxy the lowest.  Parodi \& Bingelli (2003) noted a slight but significant trend in this direction in their study of light from star-forming complexes.  In our case, the small dimension of UGC~4483 may be linked to the fact that many of its young stars are in one compact clump that acts as a separate (not scale-free) component from the field distribution.  Unfortunately, there are not enough stars to find statistically well-determined values for $d_c(r)$ for the field and cluster individually.  

A natural extension of our models would be projected, finite three-dimensional fractals.  S\'anchez et al. (2005) find that fractals with dimensions in the range $2.3-2.7$ yield projected dimensions higher than 1.85, rather close to the Poisson value of 2.  This range of $2.3-2.7$ includes the values most often inferred from observations that are treated as slices rather than projections, and is also (not surprisingly) a range given much theoretical attention.  From their correlation integrals $N(r)$, the models of S\'anchez et al. appear to flatten on large scales in a manner similar to the projected ellipsoids.  It would be interesting to examine $d_c(r)$ for these models, to verify if they really behave so similarly to the ellipsoids.  

An advantage of using $d_c(r)$ is that, like other correlation functions, it does not introduce any special scales;  if there are special scales in the distribution, they will make themselves known as features in $d_c(r)$.  Furthermore, the calculation of $d_c(r)$ does not depend on any independent characteristic of the distribution, like a determination of the center.  The major limitations in our approach, where we use galaxies that are conducive to complete photometry but that are very small, are the range of accessible scales and the number of galaxies with images at sufficiently high resolution.  These limitations are changing.  High resolution images of dwarfs continue to be made by both HST and ground-based instruments, which will allow us to examine the clustering properties of interesting but fairly distant galaxies like I~Zw~18, as well as galaxies near enough to require a larger field of view than that provided by the HST.  Studies of nearby galaxies hold the potential to provide much deeper views of the stellar population (including lower-mass stars) and to resolve smaller scales.  Finally, detailed modeling of the effects of extinction on $d_c(r)$ will open up the interpretation of resolved stellar population in larger galaxies where dust is more prevalent.  In particular, it should be possible to come up with a reasonable range of possibilities that span projection and slicing, in order to model varying transparency in a large, dusty galaxy.  

\acknowledgments 
We thank an anonymous referee for useful comments.  This paper uses data products from the Hubble Space Telescope Archive and from the Palomar/Las Companas Imaging Atlas of Blue Compact Dwarf Galaxies.

\begin{figure}
\plotone{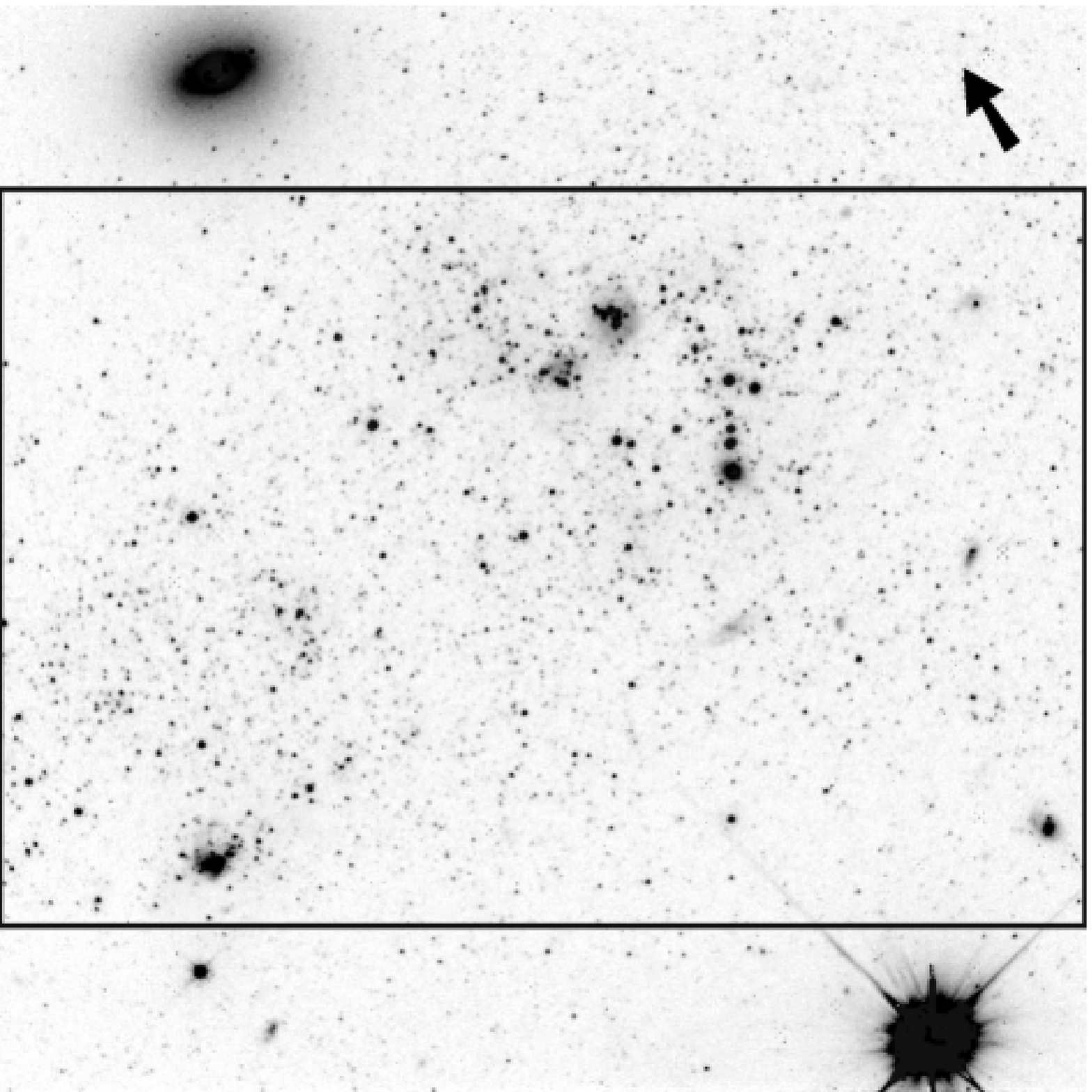}
\caption{The inner square arcminute of the galaxy UGCA~292, as seen through the HST/WFPC2 F555W filter.  The large rectangle indicates the region used to assess overall completeness and errors, and the arrow points north.} 
\label{fig1}
\end{figure}

\begin{figure}
\plotone{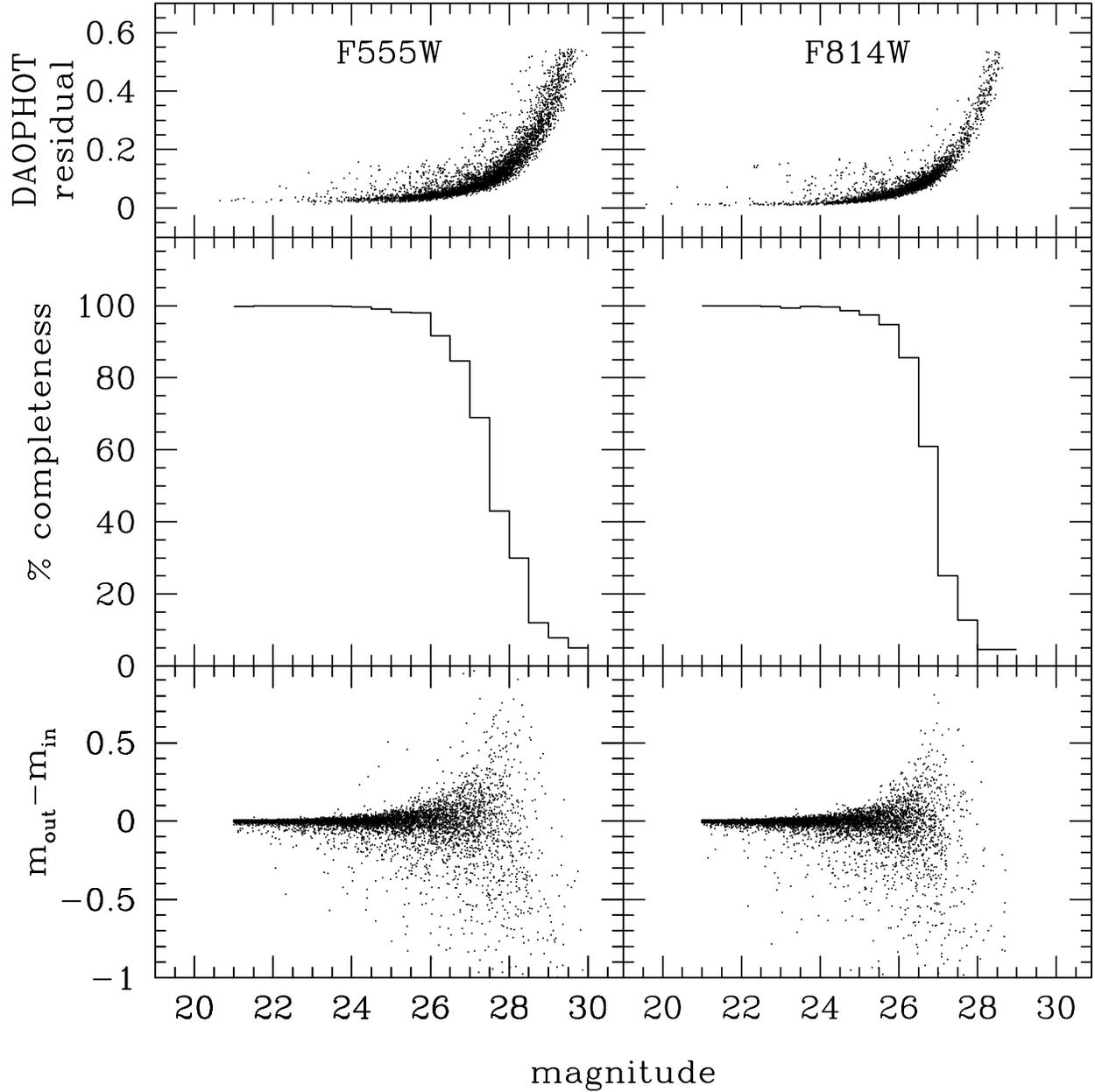}
\caption{Photometric errors and completeness for UGCA~292 in the filters F555W and F814W.  The top row shows residuals from psf fitting with DAOPHOT, while the bottom two rows show completeness and errors from artificial star tests using ADDSTAR.} 
\label{fig2}
\end{figure}

\begin{figure}
\plotone{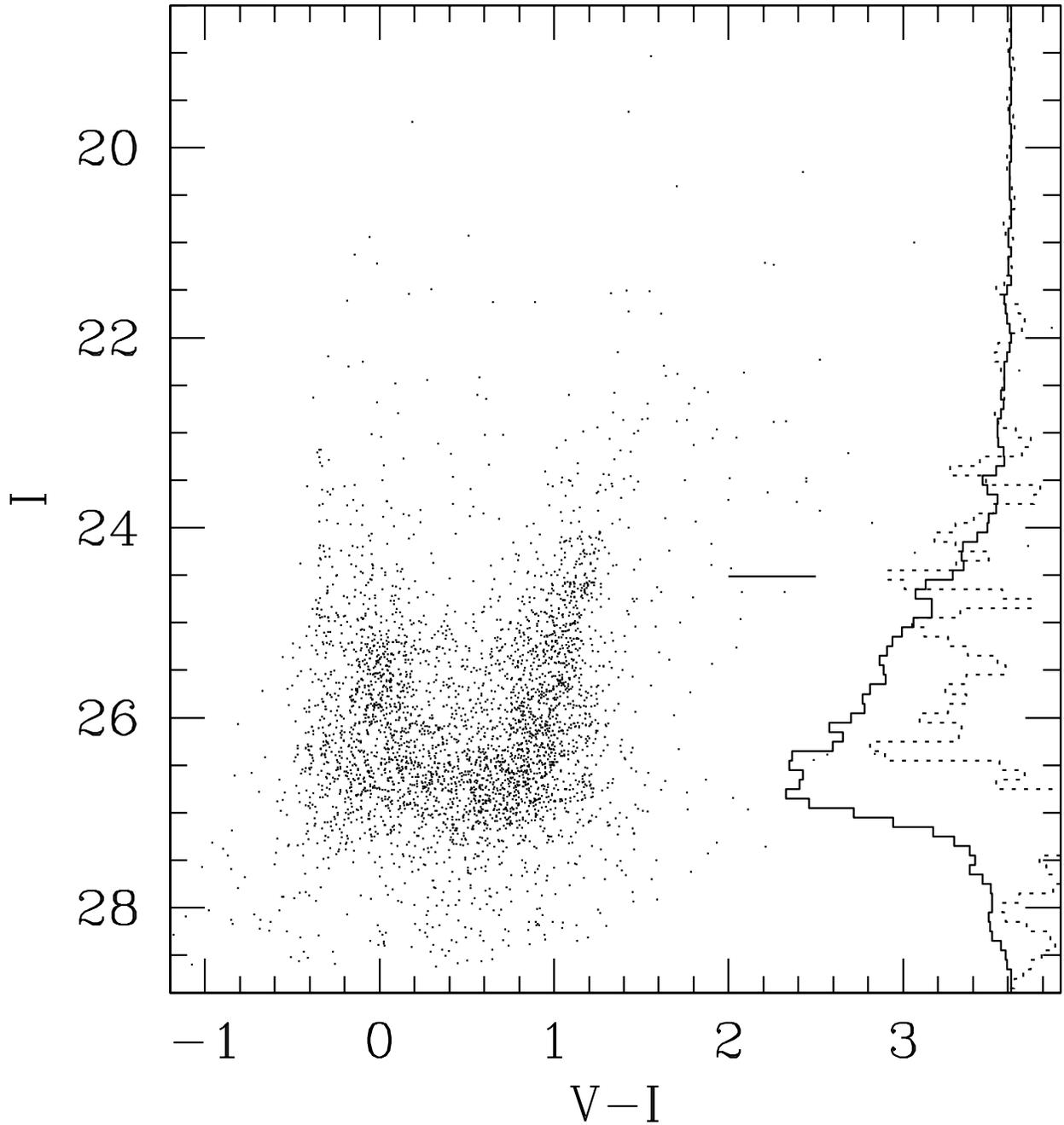}
\caption{Reddening-corrected color-magnitude diagram for UGCA~292 and, at right, the corresponding luminosity histogram (solid) and Sobel edge detector with kernel [-2, -1, 0, +1, +2] (dotted) for stars with $V-I>0.5$.  The histograms appear sideways, with magnitude scale matching the vertical scale of the CMD.  A horizontal line indicates our determination of the TRGB.} 
\label{fig3}
\end{figure}

\begin{figure}
\plotone{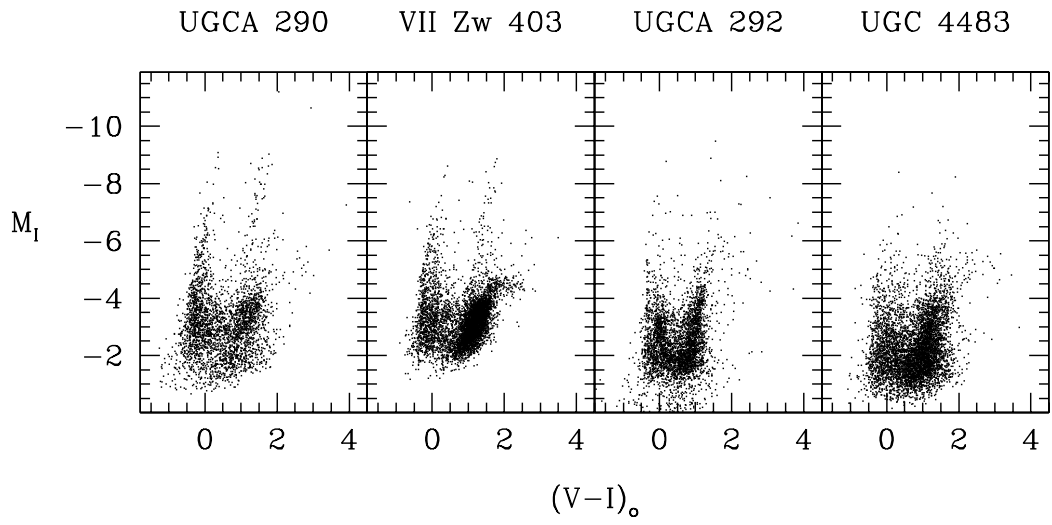}
\caption{Reddening-corrected color-magnitude diagrams of the sample galaxies, set to absolute magnitudes using the tip of the red giant branch.} 
\label{fig4}
\end{figure}

\begin{figure}
\plotone{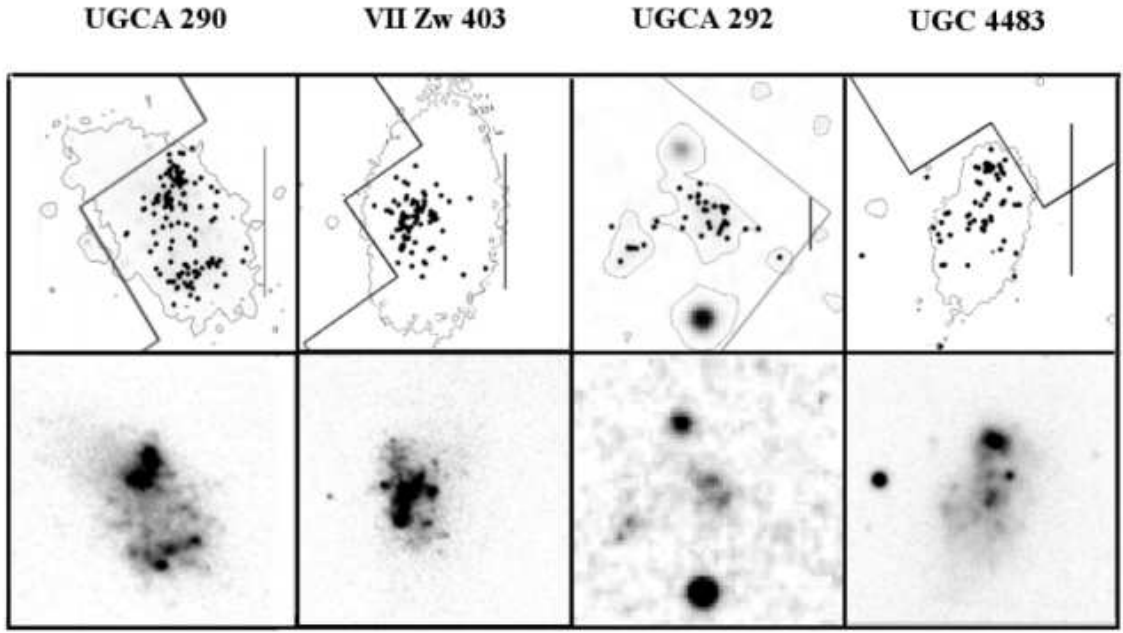}
\caption{Galaxy morphologies and coverage of the WFPC2 field of view, illustrated using ground-based images from the Palomar/Las Campanas Imaging Atlas of Blue Compact Dwarf Galaxies. North is up, and East is to the left.  Upper panels show the WFPC2 field of view, the positions of young stars resolved by the WFPC2, a vertical scale of 1 kpc, and an R-band surface brightness isophote of 24 mag arcsec$^{-2}$, which approximates the extent of the underlying older population.  Lower panels show B-band images with white set to sky background and black to 22 mag arsec$^{-2}$, illustrating the star-forming regions as dark clumps. For UGCA~292, which is not part of the Las Campanas Atlas, images are from the lower-resolution POSS-II survey.  These images show that the WFPC2 succeeds in covering most or all of the active star forming region in each galaxy.  
\label{fig5}}
\end{figure}

\begin{figure}
\plotone{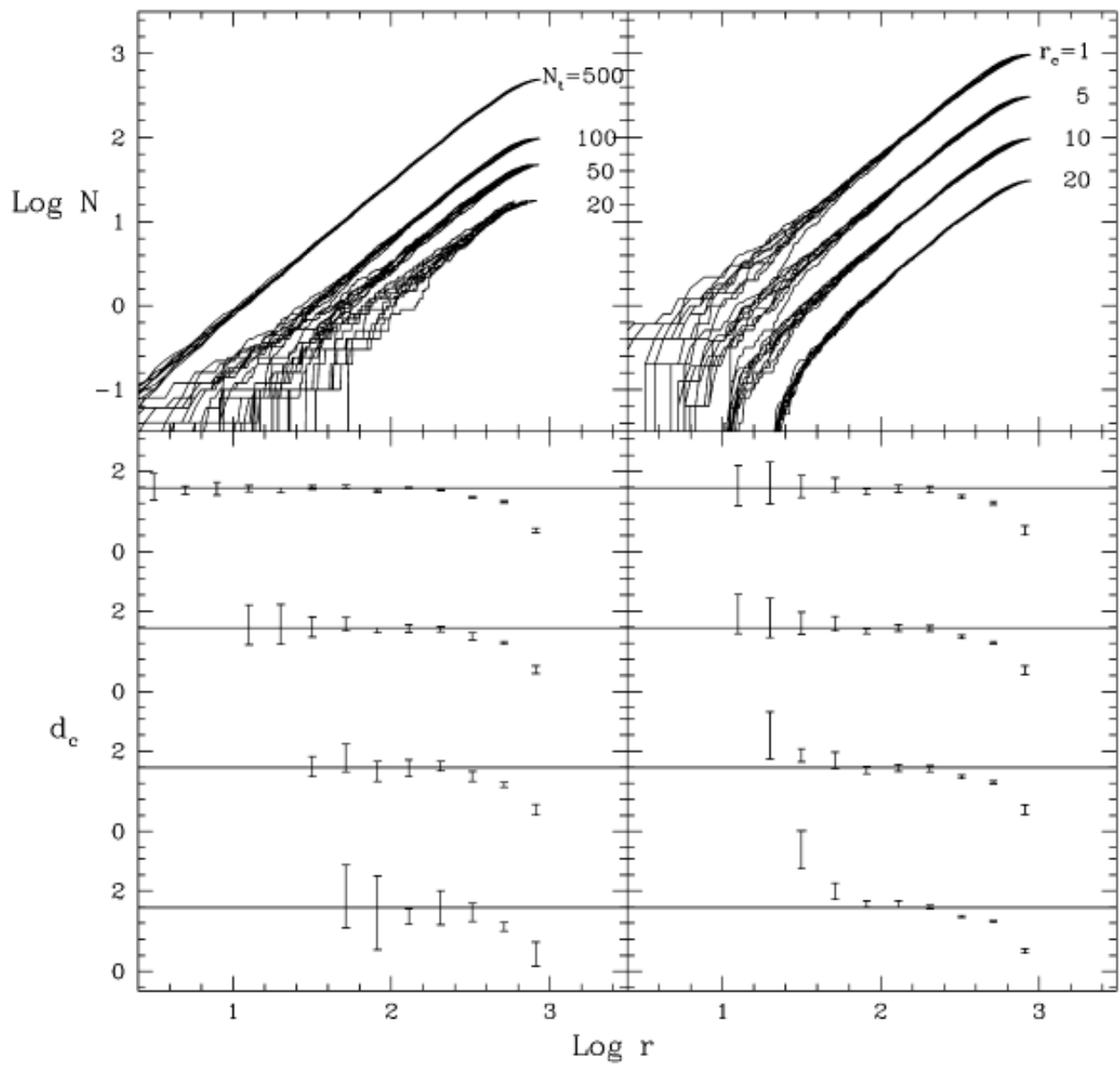}
\caption{The effect of the total number of particles $N_t$ (left) and the resolution scale $r_c$ (right) on the ability of $d_c$ to recover the dimension of a known fractal, the Sierpinski triangle.  Upper panels show the correlation integrals of ten random realization of the Sierpinski triangle for each value of $N_r$ and $r_c$, and lower panels show the corresponding values for $d_c$ in log r bins, with error bars set to the standard deviation of the random realizations. Only points for which the standard deviation does not exceed 0.7 are included. Horizontal bars show the theoretical value $d_c=1.585$.  The models with different $r_c$ all have 100 stars, but the correlation integrals are shifted vertically for clarity.}
\label{fig6}
\end{figure}

\begin{figure}
\plotone{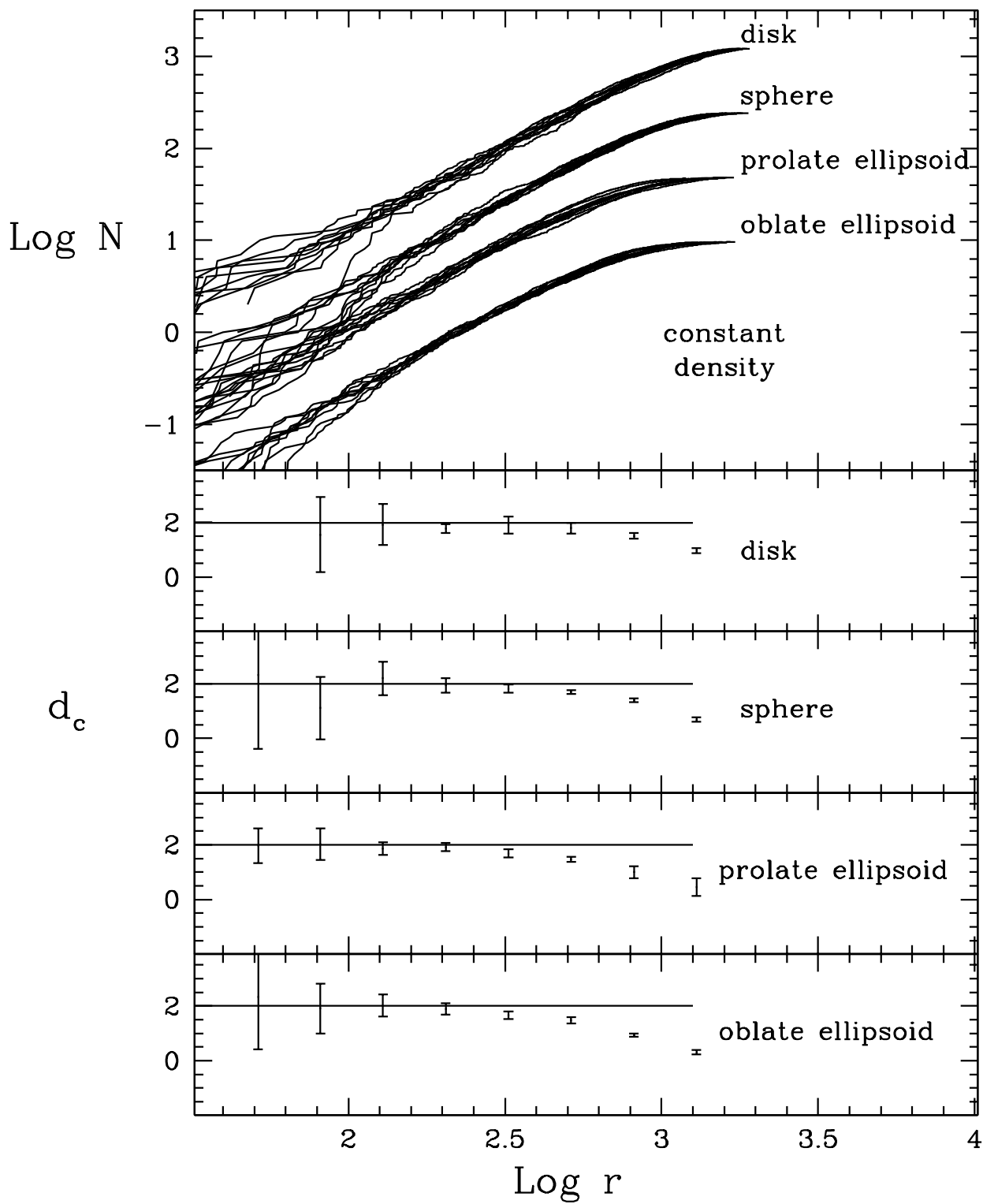}
\caption{Correlation integrals and correlation dimensions for four types of projected Poisson distributions, with semimajor axes set to 1000.  The ellipsoids have 2:1:1 and 2:2:1 axial ratios.} 
\label{fig7}
\end{figure}

\begin{figure}
\plotone{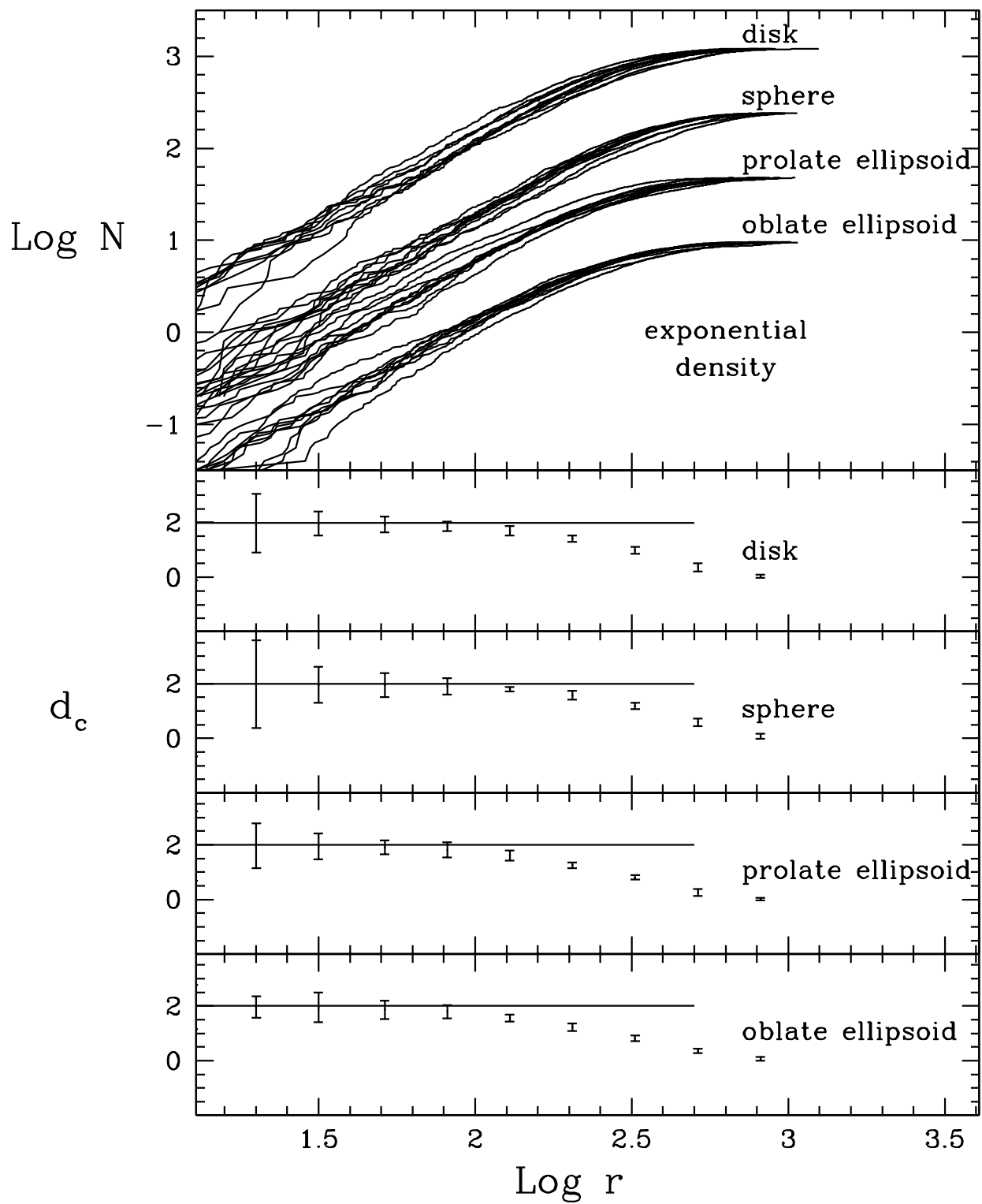}
\caption{Correlation integrals and correlation dimensions for projected distributions with exponentially decreasing density.  The format is the same as the previous figure.} 
\label{fig8}
\end{figure}

\begin{figure}
\plotone{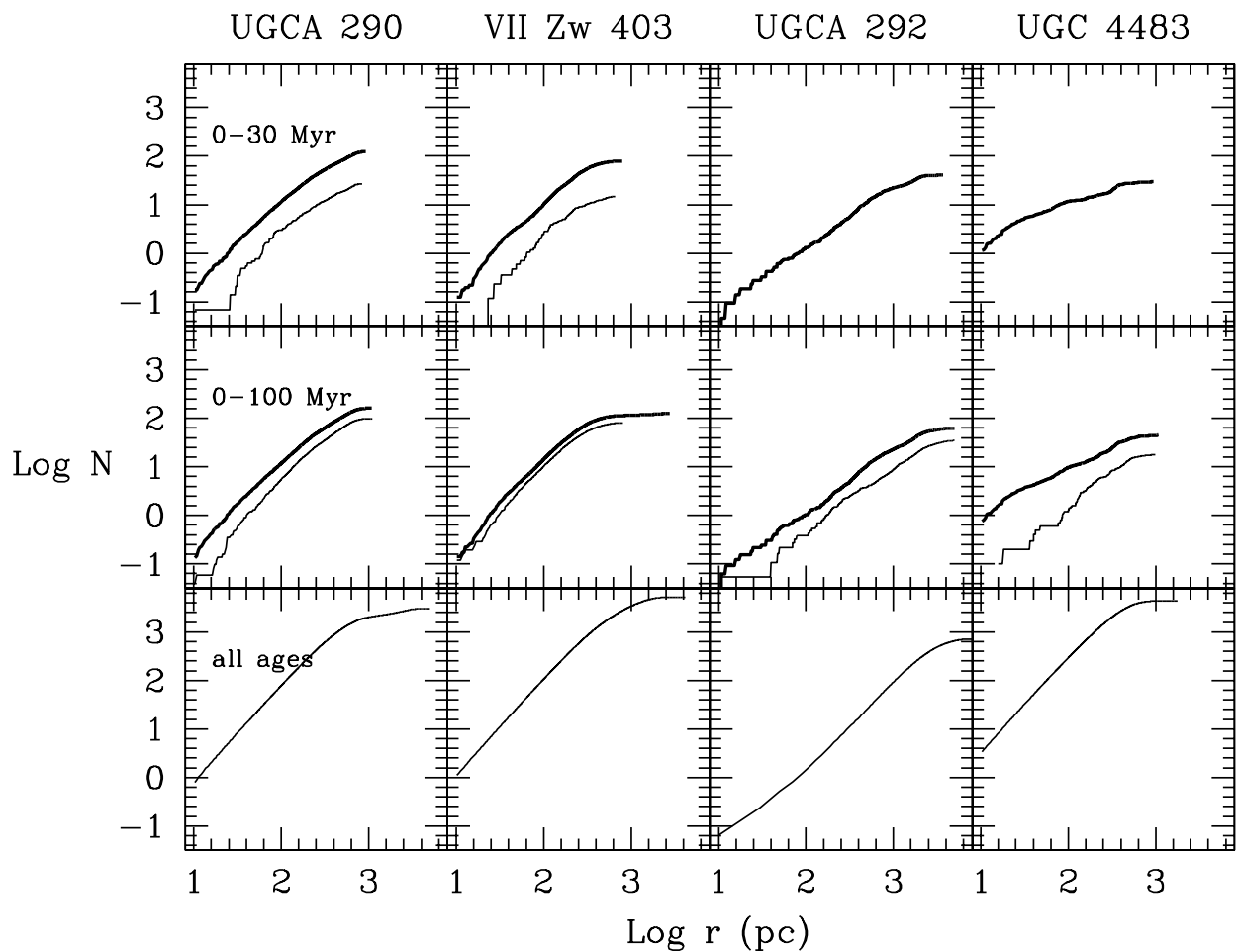}
\caption{Correlation integrals for each galaxy, for stars in the approximate age ranges $0-30$ Myr (top row) and $0-100$ Myr (middle row), and for all resolved stars (bottom row).  Bold lines include bright main sequence stars as well as supergiants while thin lines include supergiants only.  The correlation integrals for the brightest supergiants in UGC~4483 and UGCA~292 are not shown because these populations have fewer than ten stars.} 
\label{fig9}
\end{figure}

\begin{figure}
\plotone{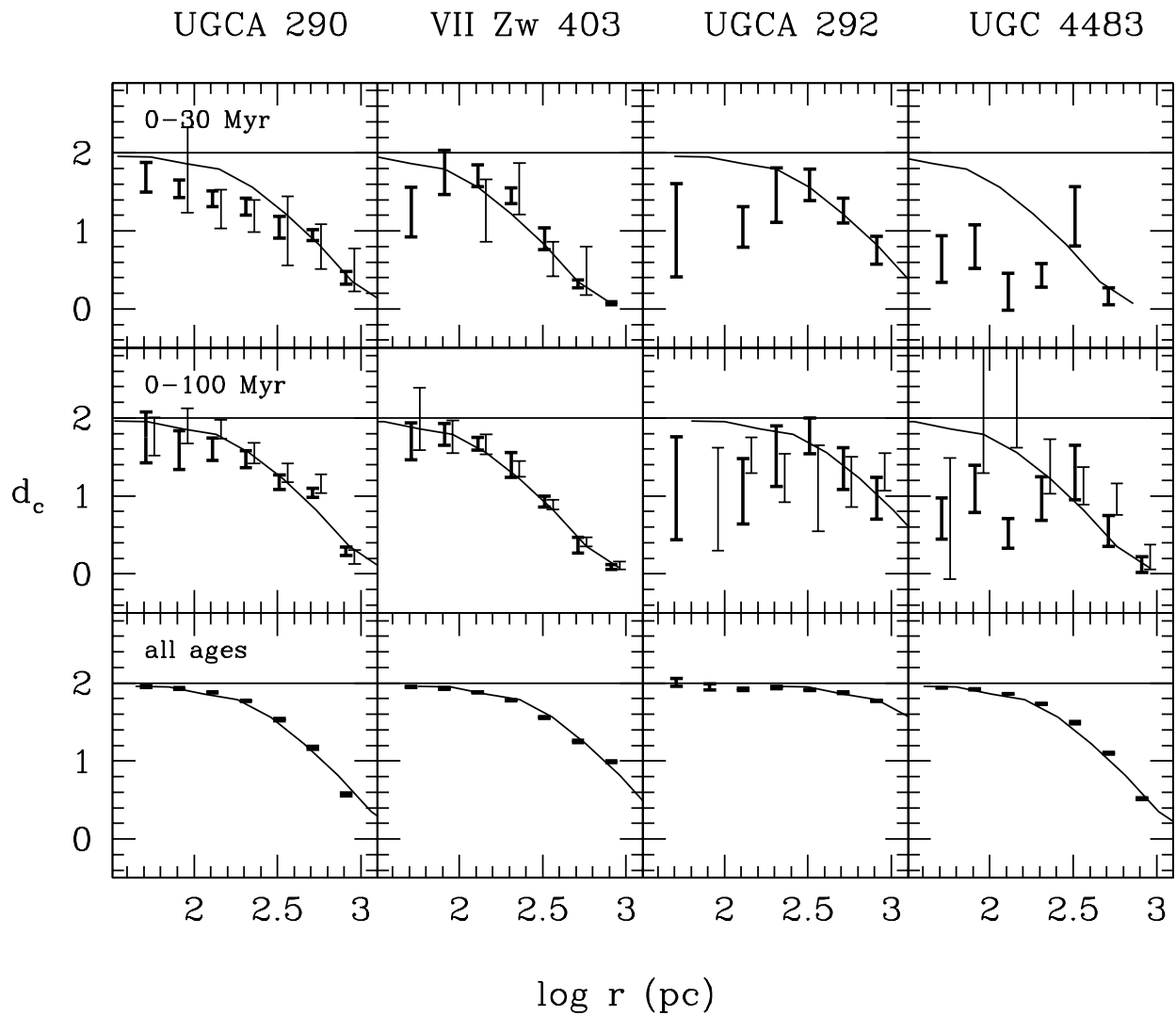}
\caption{Correlation dimensions for each galaxy, in the same format as the previous figure.  Horizontal lines indicate the value of 2.0 for a pure two-dimensional Poisson distribution, while curving lines are the results for the 2:2:1 exponential ellipsoid.  Error bars are based on bootstrap resampling, and only points with errors less than 1.0 are shown. Note that position along the horizontal axis is arbitrary for the exponential model; this curve can be shifted left or right.} 
\label{fig10}
\end{figure}

\begin{deluxetable}{lrccclll} 
\tabletypesize{\scriptsize}
\rotate
\tablecaption{Galaxy Data} 
\tablewidth{0pt}
\tablehead{
\colhead{Galaxy}&\colhead{$M_B$}&\colhead{Distance}&\colhead{12+log(O/H)}&\colhead{Proposal ID} &\colhead{Filter}&\colhead{Exposure time}&\colhead{References}\\
&&\colhead{(Mpc)}&&&&\colhead{(s)} 
} 
\startdata 
UGCA 290  	&$-13.46$	&6.7  &7.80  &8122	&F555W 	&7800		&Crone et al. 2002 \\
						&	  			& 		& 		 & 		  &F814W	&7800		& \\
VII Zw 403	&$-14.30$ &4.4  &7.73  &6276	&F555W	&4200		&Lynds et al. 1998, Schulte-Ladbeck et al 1999\\ 
						&			    &     & 		 &      &F814W	&4200		&	\\ 
UGCA 292		&$-11.43$ &5.0  &7.32  &9044	&F555W	&13000	& \\
						&		  		&			&			 &      &F814W	&26000	& \\ 
UGC 4833		&$-12.38$ &3.5  &7.56  &8769	&F555W	&9500		&Izotov \& Thuan 2002 \\
						&			    &     &    	 &			&F814W	&6900		& \\
\enddata 
\tablecomments{$M_B$ magnitudes are from Gil de Paz, Madore, \& Pevinova (2003), except that for UGCA~292, which is from van Zee (2000).  All distances are from the current study and are based on the tip of the red giant branch.  Abundances are from van Zee \& Haynes (2006), except that for VII~Zw~403, which is from Izotov, Thuan, \& Lipovetsky (1997).  References give published HST/WFPC2 photometry based on these particular data sets.}
\end{deluxetable}

\end{document}